\documentclass{article} 
\usepackage{iclr2024_conference,times}

\usepackage{amsmath,amsfonts,bm}

\def\eqref#1{equation~\ref{#1}}

\def\1{\bm{1}}

\DeclareMathAlphabet{\mathsfit}{\encodingdefault}{\sfdefault}{m}{sl}
\SetMathAlphabet{\mathsfit}{bold}{\encodingdefault}{\sfdefault}{bx}{n}

\usepackage{hyperref}
\usepackage{url}
\usepackage{comment}
\usepackage{caption}
\usepackage{multirow}
\usepackage{graphicx}
\usepackage[margin=0.95in]{geometry} 
\usepackage{cleveref}
\usepackage{glossaries}
\usepackage{booktabs}
\usepackage{gensymb}

\newacronym{edsr}{EDSR}{enhanced deep super-resolution}
\newacronym{rcan}{RCAN}{residual deep channel attention network}
\newacronym{espcn}{ESPCN}{efficient sub-pixel convolutional network}
\newacronym{ddim}{DDIM}{denoising diffusion implicit model}
\newacronym{psnr}{PSNR}{peak-signal-to-noise ratio}
\newacronym{ssim}{SSIM}{structural similarity index measure}
\newacronym{rapsd}{RAPSD}{radially averaged power spectral density}
\newacronym{melr}{MELR}{mean energy log ratio}
\newacronym{kde}{KDE}{kernel density estimation}
\newacronym{mae}{MAE}{mean average error}

\title{Wind Power Assessment based on Super-Resolution and Downscaling - A Comparison of Deep Learning Methods} 

\author{Luca Schmidt\\
Cluster of Excellence Machine Learning\\
University of Tübingen \\
Tübingen, Germany \\
\texttt{luca.schmidt@uni-tuebingen.de}
\And
Nicole Ludwig\\
Cluster of Excellence Machine Learning\\
University of Tübingen \\
Tübingen, Germany \\
\texttt{nicole.ludwig@uni-tuebingen.de} 
}
\iclrfinalcopy 
\begin{document}


\maketitle

\begin{abstract}

The efficient placement of wind turbines relies on accurate local wind speed forecasts. Climate projections provide valuable insight into long-term wind speed conditions, yet their spatial data resolution is typically insufficient for precise wind power forecasts. Deep learning methods, particularly models developed for image super-resolution, offer a promising solution to bridge this scale gap by increasing the spatial resolution of climate models. In this paper, we compare the performance of various deep learning models on two distinct tasks: \emph{super-resolution}, where we map artificially coarsened ERA5 data to its native resolution, and \emph{downscaling}, where we map native ERA5 to high-resolution COSMO-REA6 data. We evaluate the models on their downstream application in forecasting long-term wind power, emphasizing the impact of spatial wind speed resolution on wind power estimates. Our findings highlight the importance of aligning models and evaluation metrics with their specific downstream applications. We show that a diffusion model outperforms other models for estimating the wind power potential by better preserving the wind speeds' distributional and physical properties.
\end{abstract}

\section{Introduction}
Wind turbines are a key component in the transition to sustainable energy, and their success is crucially dependent on their optimal placement. An accurate understanding of local wind speed and potential changes due to climate change is essential for determining optimal turbine locations \citep{hu2023downscaling, devis2018should, tobin2016climate}.
\Citet{jung2022influence} highlight that the spatial data resolution of wind speed considerably affects the estimation of wind energy potential. Given the non-linear relationship between wind power and wind speed, even minor inaccuracies in wind speed estimates can cause substantial errors in long-term wind power forecasts. Therefore, highly localized wind speed forecasts are essential for identifying suitable sites for wind farms. 

Climate change influences wind speeds by altering the geographic distribution and variability of wind resources \citep{pryor2010climate}. As a result, areas currently viable for wind farms could see changes in wind conditions and, at the same time, new sites could become more favorable for wind energy production \citep{devis2018should}. Although long-term climate projections are typically used to evaluate future wind conditions, their coarse resolution limits their accuracy in evaluating future wind energy potential \citep{effenberger2023mind, tobin2015assessing}.

Various downscaling techniques have been developed to bridge this scale gap. \emph{(Empirical-)statistical downscaling} methods extract statistical relationships between coarse-resolution and fine-resolution climate phenomena, translating low-resolution data into localized high-resolution data. In essence, statistical downscaling learns a mapping from low-resolution to high-resolution data. Unlike classical supervised machine learning methods, samples from coarse and more fine-grained climate models have no inherent pairing. In computer vision, a problem setting that is closely related is image super-resolution. \emph{(Image) super-resolution} aims to recover a high-resolution image from its low-resolution counterpart, essentially learning the inverse of a downsampling operator \citep{wang2020deep}. Super-resolution is an ill-posed inverse problem, as multiple high-resolution images can correspond to a single low-resolution image. Existing super-resolution methods address this by imposing additional constraints to regularize the super-resolution problem, exploiting local correlations in space and time \citep{caballero2017real}.

The application of super-resolution methods has become increasingly prominent in climate downscaling. By interpreting gridded weather data as images, the image super-resolution techniques can also be applied to wind data from climate \citep{stengel2020adversarial} and weather models \citep{kurinchi2021wisosuper, miralles2022downscaling, hohlein2020comparative, yang2022statistical, hu2023downscaling}. Most research focuses on standard convolutional neural network (CNN) architectures with fully connected layers \citep{hohlein2020comparative, yang2022statistical} or generative adversarial networks (GAN) to horizontally increase the resolution of either wind velocity fields \citep{miralles2022downscaling, stengel2020adversarial, kurinchi2021wisosuper} or wind speed fields \citep{kumar2023windsr}. However, a comprehensive comparison of these methods is still lacking. While \citet{kurinchi2021wisosuper} compare some of the existing deep learning methods on a wind speed super-resolution task, their evaluation does not include a diffusion model, which \citet{merizzi2024wind} found to be the best approach for their data. 
 
In this study, we compare existing models from various model classes and evaluate their performance in the downstream task of long-term wind power estimation. By using independent high- and low-resolution models, we add complexity to the downscaling task. Specifically, we compare key super-resolution models using ERA5 reanalysis wind speed data in two tasks: a super-resolution task, where we map artificially coarsened ERA5 data back to its native resolution, and a downscaling task, where we map native ERA5 data to the higher-resolution COSMO-REA6. This dual-task analysis shows that models performing well on one task do not necessarily generalize to another. Furthermore, our evaluation highlights the importance of preserving the wind speed distribution over time for accurate long-term wind power estimation, given the non-linear relationship between wind speed and wind power, where wind speed percentiles contribute unevenly to wind power generation.

The remainder of the paper is structured as follows: \Cref{sec:methods} introduces the deep learning methods we compare in this paper. \Cref{sec:experiments} describes the two tasks with their respective data before evaluating their performance in \Cref{sec:results}.

\section{Methods} \label{sec:methods}
The computer vision community has proposed numerous single and multiple-image super-resolution methods. We select the four most promising approaches across different model classes, namely the \gls{edsr}, \gls{rcan}, \gls{espcn}, and \gls{ddim}. These methods include purely CNN-based approaches (\gls{edsr} and \gls{espcn}), an attention-based model (\gls{rcan}), and a diffusion model (\gls{ddim}). This selection aims to encompass a range of models with different architectures and inductive biases, all of which have shown promise in addressing the challenge of wind speed downscaling. 

We adjust all models to be compatible with our data and a $4\times$ downscaling factor. In each model, the velocity components, $u$ and $v$, are treated as separate inputs and outputs rather than as a combined single input representing wind speed. This approach emerged from preliminary experiments where improved accuracy was observed when treating the components individually. Each component is processed as a channel, conceptually similar to the handling of RGB channels in color images. We then compare the models against a bicubic interpolation baseline. Below, we concisely describe each model and the wind power estimation.

\subsection{Enhanced Deep Super-Resolution Network}

\begin{figure}[h!]
  \centering
  \fbox{
    \includegraphics[width=.5\textwidth]{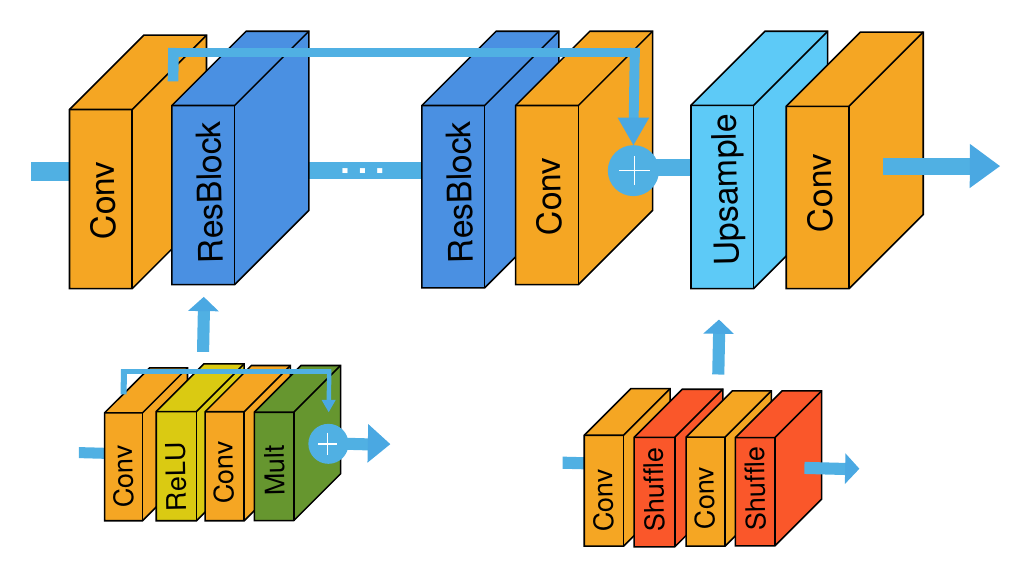}}
    \caption{The EDSR model architecture based on \cite{lim2017enhanced}}
  \label{fig:edsr_architecture}
\end{figure}

The \gls{edsr} network \citep{lim2017enhanced} is a leading CNN-based model, consistently outperforming other models in both traditional image super-resolution and wind speed super-resolution tasks \citep{zhang2018image, kurinchi2021wisosuper}. \gls{edsr} has also outperformed leading GAN models in pixel-based metrics \citep{kurinchi2021wisosuper}.

As seen in \Cref{fig:edsr_architecture}, the network consists of several convolutional layers with residual blocks and global and local skip connections. Unlike other CNN-based super-resolution methods, the \gls{edsr} has a larger model size while omitting redundant modules from conventional residual networks. The optimised training procedure with residual scaling and the end-to-end learnable sub-pixel layers used for upsampling at the network's tail make the model more computationally efficient than other approaches. We select a base model with a depth of 32 residual blocks and a width of 64 feature channels for our tasks. 

\subsection{Residual Deep Channel Attention Network}

\begin{figure}[h!]
  \centering
  \fbox{
    \includegraphics[width=.5\textwidth]{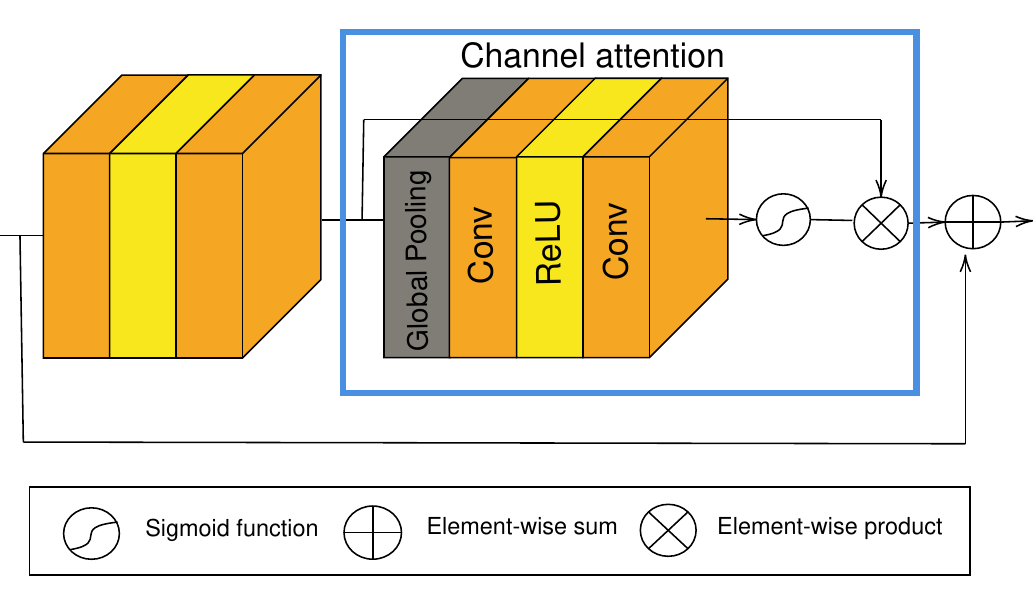}}
  \caption{Residual channel attention block based on \cite{zhang2018image}.}
  \label{fig:rcan_model}
\end{figure}

The \gls{rcan} \citep{zhang2018image} is similar to the previous \gls{edsr} model in architecture, evaluation, and training. However, it introduces an additional inductive bias through an attention mechanism. This attention mechanism allows the model to direct processing resources to the most informative components of the input, potentially leveraging local correlations more effectively.

Specifically, the \gls{rcan} incorporates a residual-in-residual structure and a channel attention mechanism, as illustrated in \Cref{fig:rcan_model}. The residual-in-residual groups consist of residual blocks with long skip connections designed to help the network focus on high-frequency information. The channel attention mechanism allows the model to consider dependencies among feature channels. We select a base model of 10 residual groups containing 20 residual channel attention blocks with 64 feature channels.

\subsection{Multi-frame Efficient Sub-pixel Convolutional Network}

\begin{figure}[h!]
  \centering
  \fbox{
    \includegraphics[width=.5\textwidth]{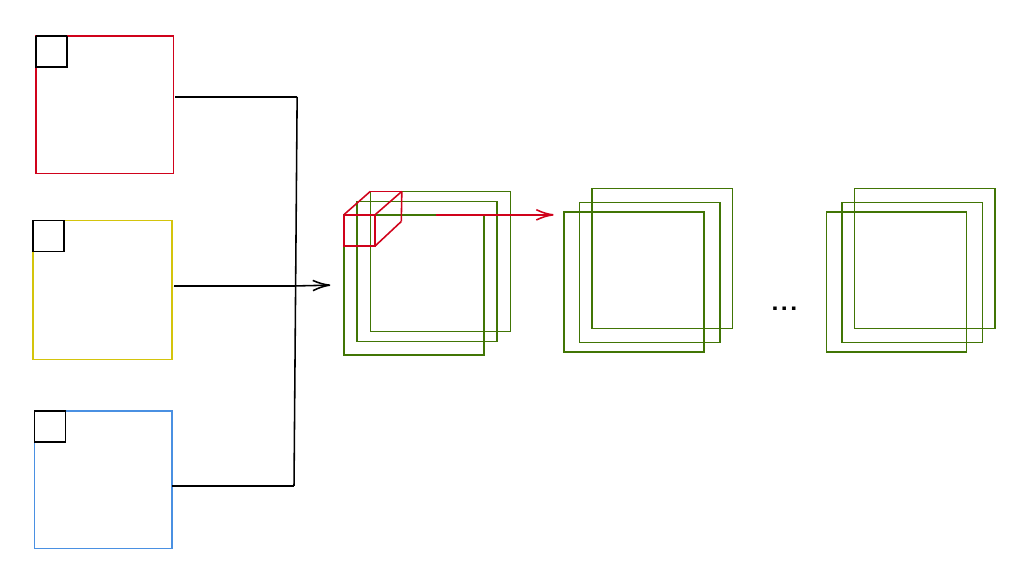}}
    \caption{Early fusion approach in multi-frame ESPCN based on \cite{caballero2017real}.}
  \label{fig:espcn_model}
\end{figure}

The two previous networks employ a one-to-one mapping of a single image, which may lead to poor image reconstruction due to the limited information in the low-resolution input image \citep{arefin2020multi}. As a result, super-resolved images may contain hallucinated details, i.e., details with no apparent relation to their original image, which occur from training bias. A solution is to increase the amount of input information by using a sequence of consecutive low-resolution images to reconstruct the high-resolution image. This approach further constrains the ill-posed super-resolution problem by exploiting temporal correlations. The multi-frame version of the \gls{espcn} introduced by \citep{caballero2017real} is a spatio-temporal model based on this idea and has been used to enhance the resolution of satellite imagery and wind speed fields \citep{liu2022video, merizzi2024wind}.

The model uses temporal fusion strategies and a motion compensation mechanism. Unlike the previous models, the \gls{espcn} parameters are optimized using an L2 instead of the L1 loss function. We select a shallow network with seven layers and use three consecutive low-resolution frames to super-resolve the middle frame. We believe that temporal dependencies beyond this number are too complex to learn. The network's temporal depth equals three, matching the number of input images. This collapses all temporal information in the first layer, corresponding to adopting an early fusion strategy (as visualized in \Cref{fig:espcn_model}). Since all the frames are perfectly aligned, there is no need to include the motion compensation mechanism. As this model does not handle multi-channel input well, we train a separate model for each wind velocity component.

\subsection{Denoising Diffusion Implicit Model}

\begin{figure}[h!]
  \centering
  \fbox{
    \includegraphics[width=14cm]{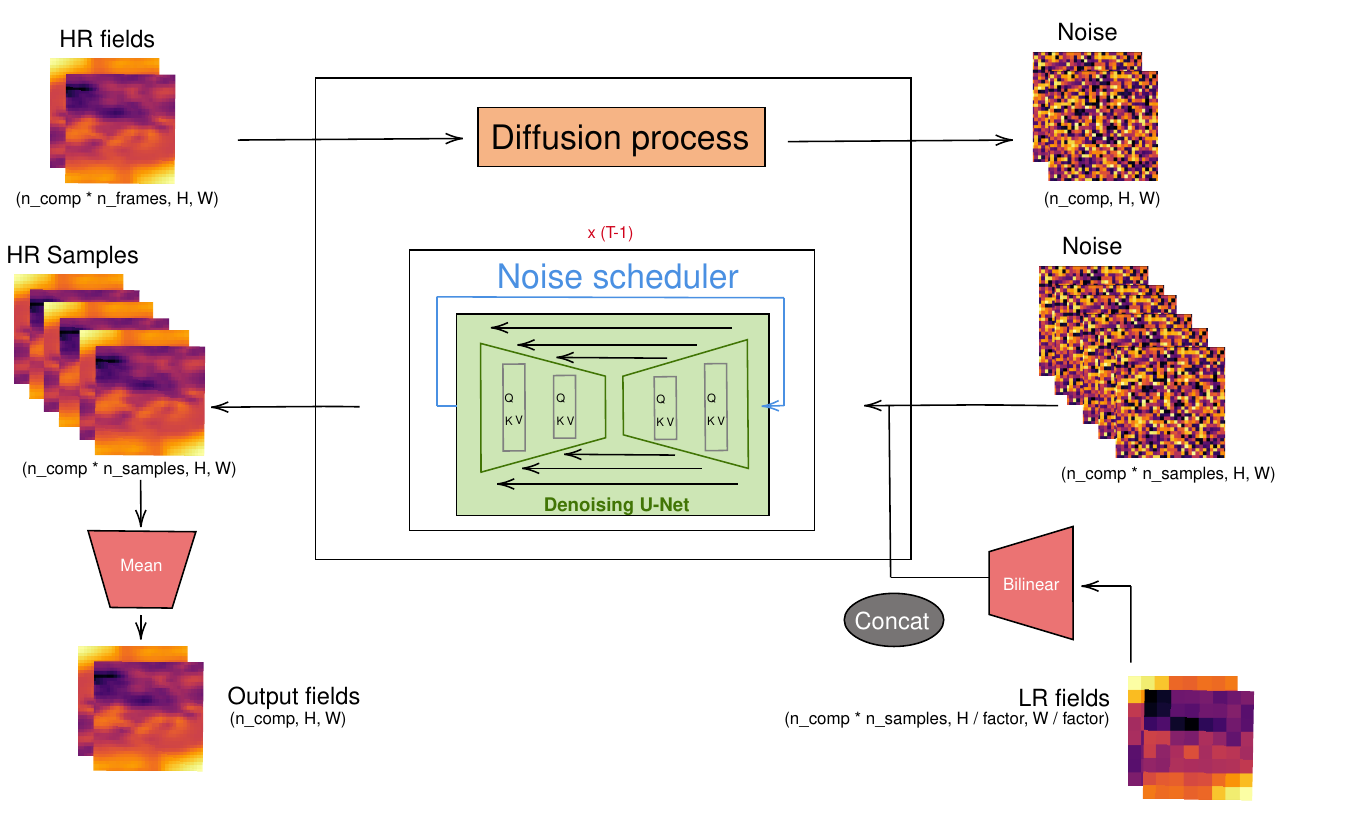}}
    \caption{Super-resolution diffusion framework based on \cite{rombach2022high} and \cite{merizzi2024wind}.}
  \label{fig:ddim_model}
\end{figure}

Diffusion models are probabilistic generative models that have demonstrated impressive results in modeling complex and high-dimensional data distributions. These models have been successfully applied to conditional and unconditional image synthesis tasks \citep{rombach2022high}. Unlike GANs, diffusion models are likelihood-based and do not encounter issues such as mode collapse or training instabilities \citep{saharia2022image}. Following the approach by \citet{merizzi2024wind}, we adopt the denoising diffusion implicit model (DDIM) proposed by \citet{song2020denoising}. DDIM is a generalization over traditional denoising diffusion probabilistic models, allowing for more efficient sample generation using non-Markovian forward processes. 

Our generative framework uses a time-conditional U-Net backbone with residual blocks and self-attention layers as suggested by \citet{rombach2022high} and illustrated in \Cref{fig:ddim_model}. The conditioning is performed by bi-linearly upsampling the low-resolution images and concatenating them to the U-Net's input via the channel dimension. 
\citet{schmidt2024benchmarking} found that using multiple frames enhances accuracy and that generating multiple samples and averaging their predictions outperforms using a single sample.
Thus, we train our model with a sequence of three high-resolution and low-resolution images of the $u$ and $v$ fields to generate samples of the middle $u$ and $v$ component fields. For each low-resolution input, we generate 15 samples and report the average performance across all samples. Similar to \citet{merizzi2024wind}, we speed up inference time by using much fewer (20) sampling steps. 

\subsection{Modeling Wind Speeds and Wind Power Generation}
The output of all the introduced models consists of two wind velocity components. These wind velocity values are rescaled to their original data range before being transformed back to wind speed to estimate wind speed and power. We estimate the wind speed distribution at grid-point locations using \gls{kde} with a Gaussian kernel and a bandwidth determined by Scott's rule \citep{scott2015multivariate}. 

Our primary objective is to analyze how the downscaled wind fields relate to wind power forecasts. Therefore, we transform the wind speed fields into wind power by assuming a wind turbine-specific power curve, which captures the relationship between wind speed at hub height and wind power \citep{lydia2014comprehensive}. For our dataset, which covers a weather domain over Germany (for more details, see \Cref{sec:experiments}), we select the Enercon E92/2350 power curve, corresponding to a wind turbine at 98 m height. Enercon is a historically dominant manufacturer of wind turbines in Germany and held the largest share in the installation of new wind turbines during the first quarter of 2024 \citep{quentin2024turbine}. Given the extensive deployment of Enercon wind turbines across Germany, they provide a reasonable model for onshore wind turbines in this region. To compute wind power values, we select random grid-point locations from the wind speed fields and apply the corresponding time series of wind speed values to the power curve model. 

\subsection{Metrics}
As we are primarily interested in wind power estimation, we evaluate the models using pixel-based metrics and metrics that assess the marginal aspects of the downscaled wind fields, i.e. statistical characteristics of the wind speed distribution over time \citep{maraun2015value}. These metrics include the \gls{psnr}, \gls{ssim} and \gls{mae} for grid-point level performance, the \gls{melr} for the preservation of physical properties, and the Wasserstein-1 distance for the temporal distribution of wind speed at specific grid-point locations.

For evaluating grid-point level performance, we rely on common image processing metrics widely used to measure the reconstruction of natural images. The \gls{psnr} measures the ratio between the peak signal strength $L$ (the maximum pixel value in the image) and the Mean Squared Error, indicating the amount of noise present in the signal \citep{wang2020deep}. The \gls{psnr} is conventionally expressed as a logarithmic quantity for interpretation in decibel units

\begin{equation}
        \mathrm{PSNR}(f, \hat{f}) = 10 \cdot \log_{10} \left(\frac{L^{2}}{\frac{1}{N}\sum_{i=1}^{N}(f(i)-\hat{f}(i)^{2})} \right) \text{\,,}
\end{equation}

where in the climate domain $f$ is the ground truth field with $N$ grid points and $\hat{f}$ is the downscaled field. As a purely pixel-based metric, the \gls{psnr} disregards structural information, such as inter-dependencies between spatially close pixel values. Therefore, \gls{psnr} is often complemented by the \gls{ssim}, which more closely aligns with perceived quality by human perception. 

The \gls{ssim} measures the structural similarity between images based on independent comparisons of luminance, contrast and structure  \citep{wang2020deep}. Luminance, $C_{l}(f, \hat{f})$ is represented by the mean image intensity $\mu_f = \frac{1}{N} \sum_{i=1}^{N} f(i)$  and contrast, $C_{c}(f, \hat{f})$ by its standard deviation $\sigma_f = \sqrt{\frac{1}{N-1} \sum_{i=1}^{N} (f(i) - \mu_f)^2}$. The structure comparison, $C_{s}(f, \hat{f})$ measures the similarity through the covariance $\sigma_{f\hat{f}} = \frac{1}{N-1} \sum_{i=1}^{N} (f(i) - \mu_f)(\hat{f}(i) - \mu_{\hat{f}})$. The \gls{ssim} is given by

\begin{equation}
\mathrm{SSIM}(f, \hat{f}) = [C_{l}(f, \hat{f})]^{\alpha}[C_{c}(f, \hat{f})]^{\beta}[C_{s}(f, \hat{f})]^{\gamma} \text{\,,}
\end{equation}
where
\begin{align}
C_{l}(f, \hat{f}) &= \frac{2\mu_f\mu_{\hat{f}} + (k_1 L)^2}{\mu_f^2 + \mu_{\hat{f}}^2 + (k_1 L)^2} \text{\,,}\\
C_{c}(f, \hat{f}) &= \frac{2\sigma_f\sigma_{\hat{f}} + (k_2 L)^2}{\sigma_f^2 + \sigma_{\hat{f}}^2 + (k_2 L)^2} \text{\,,} \\
C_{s}(f, \hat{f}) &= \frac{\sigma_{f\hat{f}} + \frac{(k_2 L)^2}{2}}{\sigma_f \sigma_{\hat{f}} + \frac{(k_2 L)^2}{2}} \text{\,,}
\end{align}
with $k_1, k_2 \gg 1$ and parameters $\alpha$, $\beta$ and $\gamma$ which adjust their relative importance. 

Additionally, we report the MAE to measure the absolute error between the downscaled and the ground truth wind fields

\begin{equation}
   \mathrm{MAE}(f, \hat{f}) = \frac{\sum_{i=1}^{N}\lvert f(i) - \hat{f}(i)\rvert}{N} \text{\,.}
\end{equation}

The first three metrics evaluate how well the downscaled wind speed fields retain location-based and spatial characteristics, such as spatial patterns and structures. However, they do not assess the preservation of the physical properties of the data. To address this, we analyze the \gls{rapsd} and derive the \gls{melr} as a metric. The \gls{rapsd} provides insight into the spatial intermittency of a 2D field at various scales by converting the 2D-power spectrum from Cartesian to polar coordinates and averaging out the angular dependence, resulting in a 1D representation of the signal \citep{ruzanski2011scale}. This representation illustrates the relative energy distribution across different spatial scales, indicating how well the physical properties are preserved. 

The energy spectrum $E(k_{x}, k_{y})$ is a 2D function of both wavenumber and direction and is derived from a Fourier transform applied to the original 2D field $f(x,y)$ of dimension $(m, n)$ 

\begin{equation}
    P(f)  = |F(k_{x}, k_{y})|^{2} = \Bigg\lvert\frac{1}{mn} \sum_{x=-m/2}^{n/2-1} \sum_{y=-n/2}^{n/2-1} f(x, y) e^{-i2\pi \left( \frac{k_x x}{m} + \frac{k_y y}{n} \right)}\Bigg\rvert^{2} \text{\,.}
\end{equation}

The \gls{rapsd} is found by averaging the power spectrum for all directions of the same wavenumber, which is a function of the radius $k = \sqrt{k_{x}^{2} + k_{y}^{2}} \text{\,,}$ such that 

\begin{equation}
E(f_r) = \frac{1}{N_r(f_r)} \sum_{i=1}^{N_r(f_r)} P(f_{r,i}) \text{\,,}
\end{equation}

where $N_{r}(k)$ is the number of frequency samples and $\lambda = s/k$ the spatial radial wavelength with $s$ being the grid spacing in Cartesian coordinates.

From the \gls{rapsd} we derive the \gls{melr}, which serves as a scalar measure to express the overall consistency between the spectrum of the predicted wind fields and the actual ones, using

 \begin{equation}
     \mathrm{MELR} = \sum_{k} \left| \log \left( \frac{E_{\text{pred}}(k)}{E_{\text{ref}}(k)} \right) \right| \text{\,.}
 \end{equation}

\gls{melr} is derived by comparing the energy of the predicted field $E_{\text{pred}}(k)$ with the reference field $E_{\text{ref}}(k)$. The \gls{melr} definition above is applied to each predicted and reference field from the test set and averaged to obtain the metric. 

Finally, we assess the distributional properties over time using the Wasserstein-1 distance. The Wasserstein-1 distance compares the temporal distributions of wind speeds at a chosen grid-point location. It measures the similarity between two probability distributions based on the concept of \emph{optimal transport} 

\begin{equation}
    \mathrm{W}(P_{\text{pred}}, P_{\text{ref}}) = \inf_{\gamma \in \Pi(P_{\text{pred}}, P_{\text{ref}})} \mathbb{E}_{(x,y) \sim \gamma} [ \| x - y \|_1 ] \text{\,,}
\end{equation}

where $\Pi(P_{\text{pred}}, P_{\text{ref}})$ is the set of couplings, that is, probability distributions whose marginals are $P_{\text{pred}}$ and $P_{\text{ref}}$. Intuitively, it quantifies the amount of probability mass $\gamma(x, y)$ associated with a particular transport plan that transforms the distribution $P_{\text{pred}}$ into the distribution $P_{\text{ref}}$. The Wasserstein-1 distance describes the expected cost of realizing the optimal transport plan \citep{arjovsky2017wasserstein}. 



\section{Experiments} \label{sec:experiments}

\begin{figure}[h!]
    \centering
    \includegraphics[width=\textwidth]{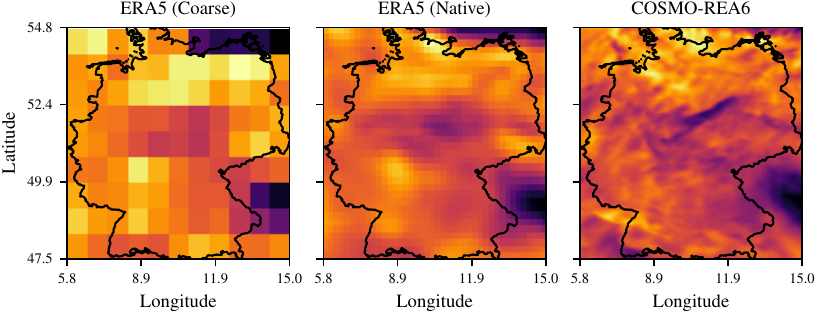}
    \caption{Example wind speed fields for the coarsened ERA5 data (left), the native ERA5 grid (middle) and the COSMO-REA6 data (right). The \emph{super-resolution} task maps the information from the left to the middle grid, while the \emph{downscaling} task maps the information from the middle to the right grid.}
    \label{fig:contours}
\end{figure}

We evaluate the performance of the deep learning methods with one experiment for each task, super-resolution and downscaling. Both experiments use weather data from Germany, specifically the $u$ and $v$ components of wind at 100 m above the surface, corresponding to a reasonable hub height for average onshore wind turbines. \Cref{fig:contours} provides an overview of the data and tasks. We map coarse ERA5 data (left) to native ERA5 data (middle) for the super-resolution task and native ERA5 data (middle) to high-resolution COSMO-REA6 data for the downscaling task.

\subsection{Super-resolution Task}
The super-resolution task aims to restore artificially coarsened ERA5 data to its original native resolution. The ERA5 reanalysis data from the European Centre for Medium-Range Weather Forecasts (ECMWF) is the gold standard for global meteorological data. It provides global reanalysis data on a regular reduced Gaussian grid with a spatial resolution of $0.25^\circ$ (equivalent to approximately 25 km for our selected region/latitudes) and hourly temporal resolution.

For this task, we extract patches of size $32 \times 32$ for the easterly and westerly wind velocities over Germany for a four-year period. The initial three years $(2017-2019)$ serve as training data, while the following year $(2020)$ is used for evaluation.
The spatio-temporal distribution of each wind velocity component is heavy-tailed. To ensure stability during training, we normalize each $u$ and $v$ component to the range $[0, 1]$ and subtract the total mean of the training set. Similarly, the test set is normalized using the values from the training set. 

To generate our low-resolution training data, we downsample the original high-resolution patches of size $32 \times 32$. Due to the grid-point representation of the ERA5 data \citep{hersbach2020era5}, it is convenient to decimate by a factor of 4, which produces the corresponding low-resolution patches of size $8 \times 8$. We believe that discarding data points this way appropriately models the degradation process. 


\subsection{Downscaling Task}
ERA5 data is the low-resolution input for the downscaling task, while the COSMO-REA6 dataset presents the high-resolution target data. COSMO-REA6 matches ERA5 in temporal resolution but offers a higher $0.055 ^{\circ}$ (approximately 6 km for our selected region/latitudes) spatial resolution. We focus on a three-year training period $(2015-2017)$ and use the subsequent year $(2018)$ for evaluation. 

To ensure spatial alignment between the two datasets, we map ERA5 to the (downsampled) COSMO-REA6 coordinates using nearest neighbor interpolation. Subsequently, we extract $136 \times 168$-sized patches over Germany from COSMO-REA6 as the high-resolution patches and patches measuring $34 \times 42$ from ERA5 as the low-resolution data. Each dataset is normalized separately to the range $[0, 1]$ using the respective min-max values, and the total mean of each dataset is subtracted, respectively. As an additional pre-processing step for the diffusion model, we bi-linearly interpolate the low-resolution patches to match the dimensions of the high-resolution patches.

\section{Results} \label{sec:results}

In this section, we evaluate all the aforementioned deep learning methods on both tasks using the specified metrics. The subsequent sections discuss our findings and outline the limitations.

\subsection{Super-resolution Task}
\begin{figure}[h!]
    \centering
    \includegraphics[width=\textwidth]{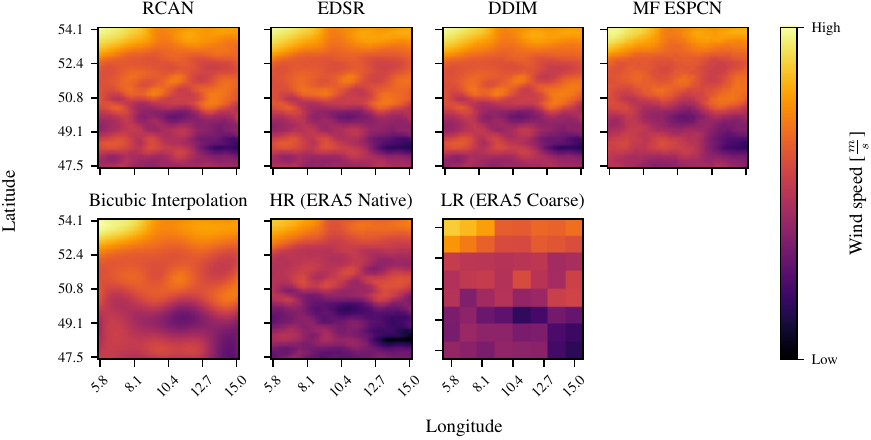}
    \caption{Exemplary wind speed fields derived from $4\times$ super-resolved $u$ and $v$ velocity fields. The input image is shown on the lower right (LR), while the target image (HR) and the baseline are shown on the lower middle and left.}
    \label{fig:sr_fields}
\end{figure}

The super-resolution task is expected to be more straightforward, as both the input and the target grid come from the same weather model. \Cref{fig:sr_fields} illustrates the wind speed fields derived from the $4\times$ super-resolved $u$ and $v$ velocity components with the input (low-resolution), target (high-resolution) and a bicubic interpolation baseline in the second row. The super-resolved fields from all models visually show a high degree of agreement with the ground truth high-resolution ERA5 field. In contrast, the bicubically interpolated field appears blurry and lacks sharp edges and fine-scale details. The super-resolution models can learn patterns not present in the LR input, which bicubic interpolation cannot capture. 

\Cref{sr_point_results} gives an overview of all metrics and further corroborates the deep learning models' ability to super-resolve the low-resolution weather data. The diffusion model slightly outperforms all other models for pixel-based metrics and representation of physical properties.

\begin{table}[h!]
      \centering
      \footnotesize
        \caption{Overview of the super-resolution task evaluation metrics. The PSNR, SSIM, MAE, and MELR are averages of all samples in the test set. The Wasserstein-1 distance is calculated at a randomly selected grid point. The best-performing model is highlighted in bold, and arrows indicate the better score direction.}
      \begin{tabular}{l l l l l l}
        \toprule
        \textbf{Model} & \textbf{PSNR} $\uparrow$ & \textbf{SSIM}$\uparrow$ & \textbf{MAE} $\downarrow$ & \textbf{MELR} $\downarrow$& \textbf{Wasserstein} $\downarrow$\\
        \midrule
        EDSR & 43.6977& 0.9773& 0.0054 & 0.1307 & 0.0349 \\
        RCAN & 43.9421& 0.9800 & 0.0053 & 0.1254 & 0.0337\\
        ESPCN & 41.8310 & 0.9644 & 0.0074 &  0.1980 & 0.1132\\
        DDIM & \textbf{45.2553}& \textbf{0.9859}& \textbf{0.0045} & \textbf{0.1077}  & \textbf{0.0299}\\
        Bicubic & 34.1660& 0.9007& 0.0178 & 0.4014  & 0.8931  \\
        \bottomrule
      \end{tabular}
  \label{sr_point_results}
\end{table}

\begin{figure}[h!]
  \centering
    \centering
    \includegraphics[width=.9\textwidth]{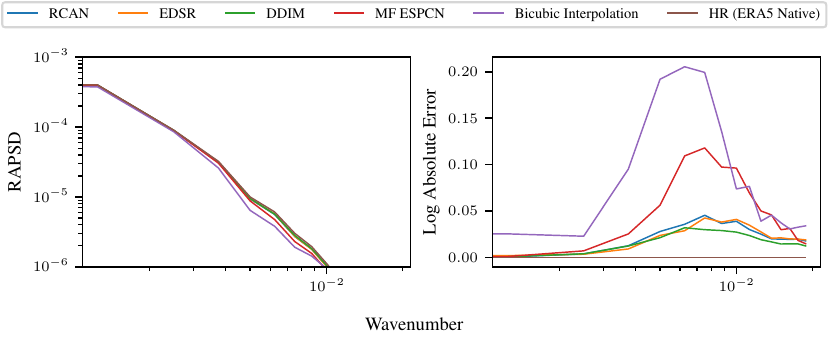}
  \caption{Left: RAPSDs of the wind speed fields as an average over the test set. Right: The absolute error of the log-transformed spectra w.r.t. the ground truth is shown to highlight the differences.}
  \label{sr_rapsd}
\end{figure}

\Cref{sr_rapsd} presents the \gls{rapsd} as a function of wavenumber, showcasing the kinetic energy transfer from larger to smaller scales, once as an average over the test set and once as an absolute error of the log-transformed spectra with respect to the ground truth. The diffusion model aligns closely with the ground truth, whereas bicubic interpolation captures high-frequency details less adequately.

We explore the implications of our findings for wind power estimation as a downstream task. Within this context, preserving the empirical distribution of wind speeds over time is more relevant than providing reliable point forecasts, as it offers a better estimate of the expected wind energy potential at a specific location \citep{devis2018should, zhou2013spatial}. 

\begin{figure}[h!]
  \centering
    \includegraphics[width=12cm]{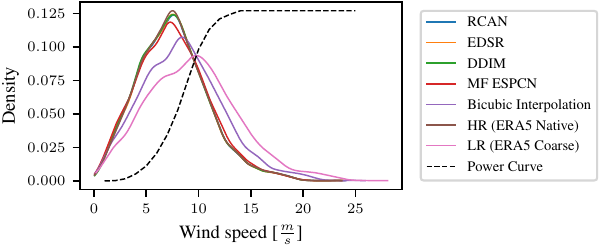}
\caption{Comparison of the super-resolved wind speed distributions (estimated with KDE) and their relation to a scaled theoretical power curve at the randomly selected grid point $(53.55\degree \mathrm{N}, 7.8\degree \mathrm{E})$} 
\label{sr_norm_ws}
\end{figure}

\Cref{sr_norm_ws} displays the estimated kernel densities at a specific grid-point location alongside the theoretical wind power curve. 
Using Wasserstein-1 distance, we assess how closely super-resolved wind speeds match the ground truth. Super-resolution models outperform bicubic interpolation and low-resolution data. This indicates that downscaling helps preserve the wind speed distribution more reliably.

\begin{figure}[h!]
  \centering
    \includegraphics[width=\textwidth]{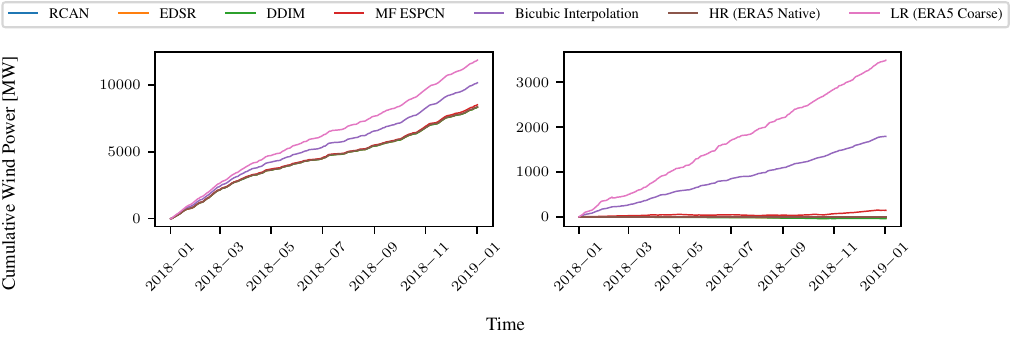}
  \caption{Cumulative wind power estimates at a selected grid-point location $(53.55\degree \mathrm{N}, 7.8\degree \mathrm{E})$ and differences w.r.t. the ground truth.}
  \label{fig:sr_cumm}
\end{figure}

The effect of small deviations between the super-resolved wind speed fields becomes more pronounced when estimating long-term wind power. \Cref{fig:sr_cumm} illustrates the cumulative wind power generated by a hypothetical turbine at a specific grid-point location over one year, revealing substantial discrepancies due to slight differences in wind speed distributions. While the low-resolution data or the bicubic interpolation tend to significantly overestimate cumulative wind power, the deep learning-based super-resolution models provide more accurate representations of local wind speed distributions. The diffusion model best matches the ground truth wind speed distribution across different locations, leading to the most accurate long-term wind power estimates.

\subsection{Downscaling Task}
\begin{figure}
    \centering
    \includegraphics[width=\textwidth]{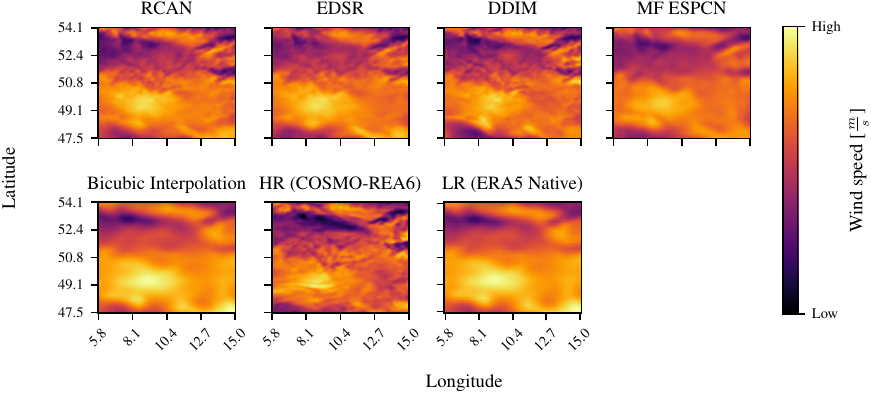} 
    \caption{Exemplary wind speed fields derived from $4\times$ downscaled $u$ and $v$ velocity fields. The input image is shown on the lower right (LR), while the target image (HR) and the baseline are shown on the lower middle and left.}
    \label{fig:dsc_fields}
\end{figure}

In the downscaling task, we examine whether our findings generalize to the more complex scenario where low- and high-resolution data originate from different weather models. The downscaling task reveals substantial qualitative differences between the downscaled wind and ground truth fields compared to the super-resolution task, as seen from the wind speed fields in \Cref{fig:dsc_fields}. Unlike the super-resolution task, the increased level of small-scale detail does not necessarily improve pixel-based metrics. 

The metrics in \Cref{tab:dsc_point_metrics} show that the multi-frame \gls{espcn} model achieves high PSNR and low MAE, indicating comparatively good performance on purely pixel-based metrics. However, it produces overly smooth wind fields, similar to bicubic interpolation. Visually, the diffusion model shows the highest level of detail and resemblance to the ground truth, but this level of detail is not reflected in the pixel-based metrics. 

\begin{table}[h!]
      \centering
      \footnotesize
    \caption{Overview of the downscaling task evaluation metrics. The PSNR, SSIM, MAE, and MELR are averages of all samples in the test set. The Wasserstein-1 distance is calculated at a randomly selected grid point.}
      \begin{tabular}{l l l l l l}
        \toprule
       \textbf{Model} & \textbf{PSNR} $\uparrow$ & \textbf{SSIM}$\uparrow$ & \textbf{MAE} $\downarrow$ & \textbf{MELR} $\downarrow$& \textbf{Wasserstein} $\downarrow$\\
        \midrule
        EDSR &33.8356 & 0.9259& 0.0182 & 0.4065  & 0.1640 \\
        RCAN & 33.6769 & 0.9295& 0.0186 & 0.3771 & 0.1243\\
        ESPCN & \textbf{34.1353}&0.9089& \textbf{0.0173} & 0.5370 & 0.5543 \\
        DDIM &33.7956 & \textbf{0.9324} & 0.0183 & \textbf{0.2623} & \textbf{0.1195}\\
        Bicubic & 31.8012& 0.9027& 0.0229 & 0.4654  & 0.7048 \\
        \bottomrule
      \end{tabular}
      \label{tab:dsc_point_metrics}
\end{table}

\begin{figure}[h!]
  \centering
 \includegraphics[width=\textwidth]{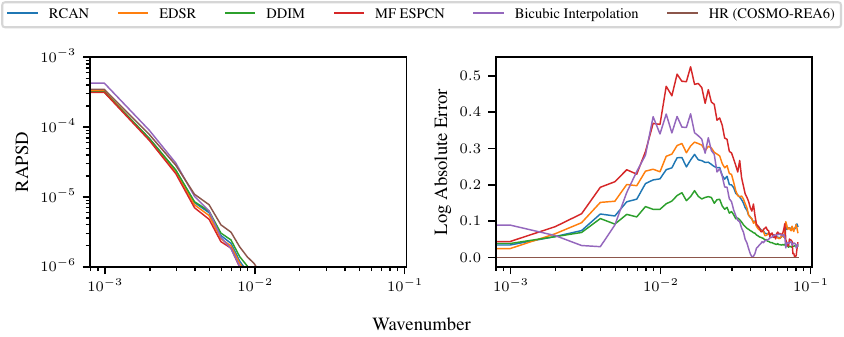}
  \caption{(a) RAPSDs of the wind speed fields as an average over the test set. (b) The absolute error of the log-transformed spectra w.r.t. the ground truth is shown to highlight the differences.}
  \label{fig:dsc_rapsd}
\end{figure}

When evaluating metrics that measure physical properties and distributions, the \gls{ddim} outperforms the other models by more accurately replicating the ground truth energy spectrum. This can also be seen in the RAPSD in \Cref{fig:dsc_rapsd}.

\begin{figure}[h!]
  \centering
    \includegraphics[width=12cm]{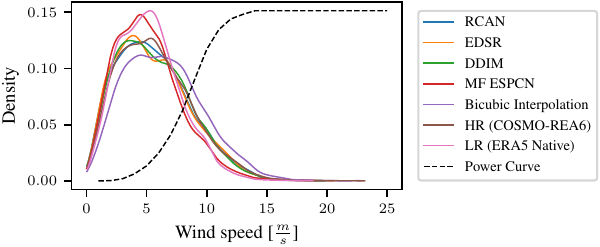}
    \caption{Comparison of the downscaled wind speed distributions (estimated with KDE) and their relation to a scaled theoretical power curve at the randomly selected grid point $(50.61\degree \mathrm{N}, 7.75\degree \mathrm{E})$.}
    \label{fig:dsc_kde_ws}
\end{figure}

\begin{figure}[h!]
    \centering
    \includegraphics[width=\textwidth]{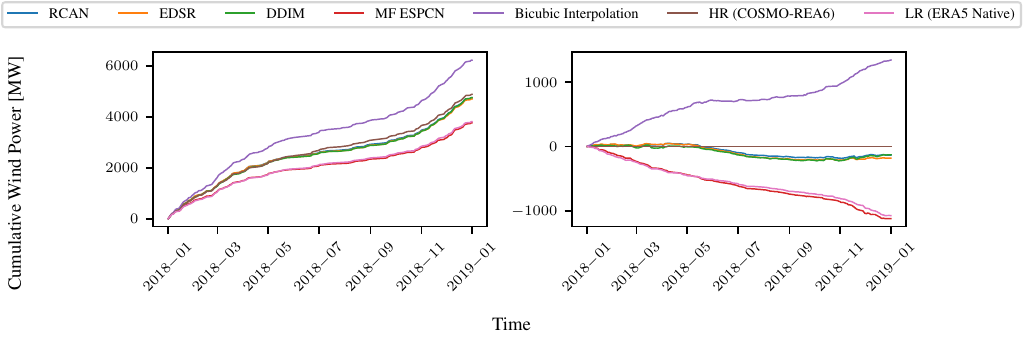}
    \caption{Left: Estimated cumulative wind power at a selected grid-point location $(50.61\degree \mathrm{N}, 7.75\degree \mathrm{E})$. Right: Difference in estimated cumulative wind power w.r.t the ground truth.}
\label{fig:dsc_cumm}
\end{figure}

In our analysis of long-term wind power estimation, we examine the differences in wind speed distributions generated by the different models (see \Cref{fig:dsc_kde_ws}) and their impact on cumulative wind power, see \Cref{fig:dsc_cumm}. These differences are more pronounced compared to the super-resolution case. Except for the multi-frame \gls{espcn} and bicubic interpolation, all models yield more accurate wind power estimates than the low-resolution data. In most cases, the diffusion model outperforms the other models in preserving the wind speed distributions over time, leading to more precise wind power estimates. 

\subsection{Discussion} \label{sec:discussion}
Our findings underscore the limitations of pixel-based metrics in capturing relevant information for our downstream application. The rankings provided by classical computer vision metrics in the downscaling task can be misleading, which becomes apparent when considering application-based metrics. Since deep learning models are typically trained using pixel-based losses, future investigations should explore their performance with probabilistic losses to better align with real-world applications.

While we use deep learning models to establish mappings between data from different weather models, our approach does not involve bias correction. Consequently, we assume that the models can learn any biases that result from the difference in properties modeled by the low- and high-resolution weather models. The discrepancy between the results of the super-resolution and downscaling task might, however, partially be caused by this lack of bias correction.

Overall, we observe that low-resolution data inadequately represents wind speeds, largely resulting in overestimations in long-term wind power estimates. This highlights the need for high-resolution wind speed information. In particular, the \glsentrylong{ddim} is the most promising model for both the super-resolution and the downscaling task of wind speed and wind power.

\section{Conclusion}
Global and regional climate models provide useful insight into future patterns of climate variables such as wind speed but are often too coarse to inform about local impact. Downscaling methods offer a promising solution to bridge the gap between coarse and high-resolution data, enabling more accurate local wind speed estimations and, consequently, better assessments of wind energy potential. This paper uses low- and high-resolution weather data and compares state-of-the-art deep learning methods on the two distinct tasks of super-resolution and downscaling wind speed. In addition, we investigate the wind power estimates resulting from the different wind speed fields. 

Our results show that the relative performance of different super-resolution models varies depending on the specific task. We show that metrics quantifying wind speed distribution over time are important, and the diffusion model, in particular, preserves the distributional aspects and physical properties of wind fields well, closely matching the ground truth cumulative energy generation. 
 
Future research should investigate the models' out-of-distribution behavior, especially on different spatial scales, time periods, and diverse geographic regions.


\subsubsection*{Acknowledgments}
This paper was funded by the Deutsche Forschungsgemeinschaft (DFG, German Research Foundation) under Germany's Excellence Strategy – EXC number 2064/1 – Project number 390727645 and the Tübingen AI Center. The authors thank the International Max Planck Research School for Intelligent Systems (IMPRS-IS) for supporting Luca Schmidt.
The authors want to thank Marvin Pförtner, Nina Effenberger, and Julia Fullenkamp for the discussions and feedback.

\bibliography{sr_ds_paper}
\bibliographystyle{iclr2024_conference}

\appendix
\section{Appendix}
You may include other additional sections here.
\end{document}